\documentclass[twocolumn,pra,aps,showpacs]{revtex4}

\usepackage{amsmath}
\usepackage{amssymb}
\usepackage[dvips]{graphicx}
\usepackage{pstricks}
\usepackage{pst-node}
\usepackage{pst-plot}

\newcommand{\qed}{\hspace*{\fill}$\square$}

 \newcommand{\Z}{\mathbf{Z}}

 \newcommand{\sset}[1]{ \{#1\} }

 \newcommand{\ket}[1]{|#1\rangle}

 \newcommand{\Dn}{{\cal H}_2^{\otimes n}}
 
 \newcommand{\Pauli}[1]{\mathbf P_{#1}}

\begin{document}

\title[Short Title]{
Topological Quantum Distillation}

\author{H. Bombin and M.A. Martin-Delgado}
\affiliation{
Departamento de F\'{\i}sica Te\'orica I, Universidad Complutense,
28040. Madrid, Spain.
}

\begin{abstract}
We construct a class of  topological quantum codes to perform
quantum entanglement distillation. These codes implement the whole
Clifford group of unitary operations in a fully topological manner
and without selective addressing of qubits. This allows us to extend
their application also to quantum teleportation, dense coding and
computation with magic states.
\end{abstract}

\pacs{03.67.-a, 03.67.Lx}

\maketitle

One of the main motivations for introducing  topological error
correction codes \cite{kitaev97}, \cite{dennis_etal02},
\cite{bravyikitaev98} in quantum information theory is to realize a
naturally protected quantum system: one that is protected from local
errors in such a way that there is no need to explicitly perform an
error syndrome measurement and a fixing procedure. Physically, this
is achieved by realizing the code space in a topologically ordered
quantum system. In such a system we have a gap to system excitations
and topological degeneracy, which cannot be lifted by any local
perturbations to the Hamiltonian. Only topologically non-trivial
errors are capable of mapping degenerate ground states one onto
another. Thus, a natural question is how to implement quantum
information protocols in a topological manner, thereby getting the
benefits provided by quantum topology.

Quantum distillation of entanglement is one of those very important
protocols in quantum information \cite{bennett_etal96}. It allows us
to purify initially mixed states with low degree of entanglement
towards maximally entangled states, which are needed in many quantum
information tasks. The most general description of entanglement
distillation protocols \cite{bennett_etal96}, \cite{qpa},
\cite{numbertheory} relies on the implementation of  unitary
operations from the Clifford group. This is the group of unitary
operators acting on a system of $n$ qubits that map the group of
Pauli operators onto itself under conjugation.

In this paper we have been able to construct quantum topological
codes that allows us to implement the Clifford group in a fully
topological manner. The Clifford group also underlies other quantum
protocols besides distillation. Thus, as a bonus, we obtain complete
topological implementations of quantum teleportation and superdense
coding. We call these topological codes
 triangular codes. In
addition, they have two virtues: 1/ there is no need for selective
addressing and 2/ there is no need for braiding quasiparticles. The
first property means that we do not have to address any particular
qubit or set of qubits in order to implement a gate. The second one
means that all we use are ground state operators, not quasiparticle
excitations.

In order to achieve these goals, we shall proceed in several stages.
First, we introduce a new class of topological quantum error
correcting codes that we call \emph{color codes}. Unlike the
original topological codes in \cite{kitaev97}, these are not based
in a homology-cohomology setting. Instead, there is an interplay
between homology and a property that we call color for visualization
purposes. This \emph{color} is not a degree of freedom but a
property emerging from the geometry of the codes. After color codes
have been presented for closed surfaces, we show how colored borders
can be introduced by doing holes in a surface. In particular we
define \emph{triangular codes}, so called because they consist of a
planar layer with three borders, one of each color. These codes
completely remove the need of selective addressing. If the lattice
of a triangular code is suitably chosen, we show that the whole
Clifford group can be performed on it. Finally, we give the
Hamiltonian that implements the desired self-correcting
capabilities. It is an exactly solvable local Hamiltonian defined on
spin-1/2 systems placed at the sites of a 2-dimensional lattice.

\begin{figure}
 \includegraphics[width=8cm]{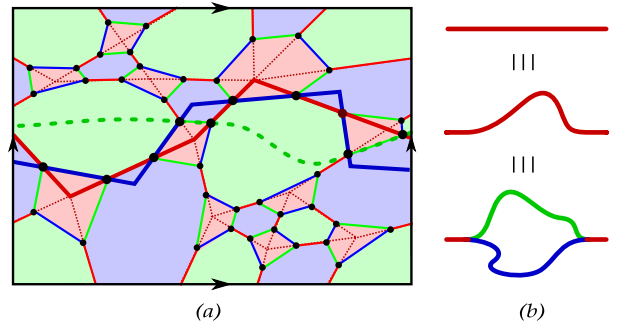}
 \caption{(a) A color code in a torus.
 Each site is a qubit and each
 plaquette a generator of the stabilizer $\mathcal S$. The dashed red
 line corresponds to the shrunk red lattice.
 The thick red and blue lines are string operators.
 They act on the sites lying on their links.
  The dotted
 green line is the string operator that results from
 the product of the red and the blue one.
  (b) There are two ways in which we can change the shape of a red string operator.
  We can either consider homologous strings only or also use the
operator equivalence \eqref{combina_colores}.}
  \label{toro_color}
\end{figure}

 A \emph{quantum error correcting code of length $n$} is a subspace $\mathcal C$ of $\Dn$,
with ${\cal H}_2$ the Hilbert space of one qubit. Let the length of
an operator on $\Dn$ be the number of qubits on which it acts
nontrivially. We say that the code $\mathcal C$  corrects $t$ errors
when it is possible to recover any of its (unknown) states after any
(unknown) error of length at most $t$ has occurred. Let
$\Pi_{\mathcal C}$ be the projector onto $\mathcal C$. We say that
$\mathcal C$ detects an operator $\mathcal O$ if $\Pi_{\mathcal
C}\mathcal O \Pi_{\mathcal C}\propto \Pi_{\mathcal C}.$ The
\emph{distance} of a code is the smallest length of a non-detectable
error. A code of distance $2t+1$ corrects $t$ errors.
 We talk about $[[n,k,d]$] codes
when referring to quantum codes of length $n$, dimension $2^k$ and
distance $d$. Such a code is said to encode $k$ \emph{logical}
qubits in $n$ \emph{physical} qubits.

 Now let $X$, $Y$ and $Z$
denote the usual Pauli matrices. A Pauli operator is any tensor
product of the form $\bigotimes_{i=1}^n \sigma_i$ with $\sigma_i \in
\sset{1,X,Y,Z}$. The closure of such operators as a group is the
\emph{Pauli group} $\Pauli n$. Given an abelian subgroup $\mathcal
S\subset\Pauli n$ not containing $-I$, an \emph{stabilizer code of
length $n$} is the subspace $\mathcal C\subset\Dn$ formed by those
vectors with eigenvalue 1 for any element of $\mathcal S$
\cite{gottesman96}, \cite{calderbank_etal}. If its length is $n$ and
$\mathcal S$ has $s$ generators, it will encode $k=n-s$ qubits. Let
$\mathcal Z$ be the centralizer of $S$ in $\Pauli n$, i.e., the set
of operators in $\Pauli n$ that commute with the elements of
$\mathcal S$. The distance of the code $\mathcal C$ is the minimal
length among the elements of $\mathcal Z$ not contained in $\mathcal
S$ up to a sign.

 Suppose that we have a 2-dimensional lattice embedded in a torus
 of arbitrary genus such that three links meet at each site and plaquettes can be
 3-colored, see Fig.~\ref{toro_color} for a example in a torus of genus one.
 We will take red, green and blue as colors ($RGB$). Notice that we can attach a color to
the links in the lattice according to the plaquettes they connect: a link
that connects two red plaquettes is red, and so on. With such an
embedding at hand we can obtain
   a \emph{color code} by choosing as generators for
$\mathcal S$ suitable plaquette operators. For each plaquette $p$
there is a pair of operators: $B_p^X$ and $B_p^Z$. Let $I$ be an
index set for the qubits in $p$'s border, then
\begin{equation}\label{operadores_borde}
B_p^{\sigma} := \bigotimes_{i\in I} \sigma_i,\qquad \sigma=X,Z.
\end{equation}
Color codes are local because \cite{kitaev97} each generator acts on
a limited number of qubits and each qubit appears in a limited
number of generators, whereas there is no limit in the code
distance, as we shall see.

We will find very useful to introduce the notion of \emph{shrunk
lattices}, one for each color. The red shrunk lattice, for example,
is obtained by placing a site at each red plaquette and connecting
them through red links, see Fig.~\ref{toro_color}. Note that each
link of a shrunk lattice corresponds to two sites in the colored
one. Note also that green and blue plaquettes correspond to the
plaquettes of the red shrunk lattice.

We classify the plaquettes according to their color into three sets,
$R$, $G$ and $B$. Observe that for $\sigma=X,Z$
\begin{equation}\label{operadores_sobrantes}
\prod_{p\in R} B_p^{\sigma}= \prod_{p\in G} B_p^{\sigma}
=\prod_{p\in B} B_p^{\sigma},
\end{equation}
hold because these products equal $\hat \sigma:=\sigma^{\otimes n}$. We
shall be using this hat notation for operators acting bitwise on the
physical qubits of the code. Equations \eqref{operadores_sobrantes}
implies that four of the generators are superfluous. We can now
calculate the number of encoded qubits using the Euler
characteristic of a surface $\chi = f+v-e$. Here $f$, $v$ and $e$
are the number of plaquettes, sites and links of any lattice on the
surface. Applying the definition to a shrunk lattice we get
\begin{equation}\label{}
k = 4-2\chi.
\end{equation}
Observe that the number of encoded qubits depends only upon the
surface, not the lattice. When the code is rephrased in terms of a
ground state in a quantum system\eqref{Hamiltoniano}, this will be
an indication of the existence of topological quantum order \cite{wenbook04}.

\begin{figure}
 \includegraphics[width=8cm]{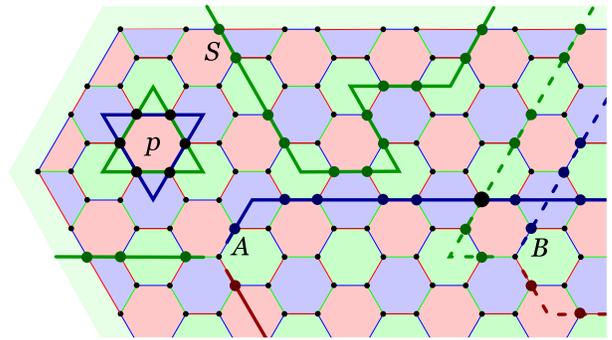}
 \caption{\emph A honeycomb lattice with a green border.
 Notice the two possible points of view for the operators of
  the plaquette $p$ as boundary paths.
  The green string $S$ is homologous to part of the border, and thus is
 equivalent to the identity.
 There is also a pair of equivalent 3-string operators, $A$ and $B$.}
 \label{borde_barion}
\end{figure}

So far we have described the Hilbert space of the logical qubits in
terms of the stabilizer. Now we want to specify the action of
logical operators on those qubits. To this end we introduce an
equivalence relation among the operators in $\mathcal Z$, which we
shall use repeatedly. We say that $A\sim B$ if $A$ and $B$ represent
the same quotient in $\mathcal Z/S$. Notice that two equivalent
operators will have the same effect in $\mathcal C$. Now we
introduce the key idea of \emph{string operators}. They can be red,
green or blue, depending on the shrunk lattice we are considering. Let
$P$ be any closed path in a shrunk lattice. We attach to it two
operators: if $P$ is a path and the qubits it contains are indexed
by $I$, we define
\begin{equation}\label{operadores_cuerda}
S_P^\sigma := \bigotimes_{i\in I} \sigma_i, \qquad \sigma = X,Z.
\end{equation}
 The
point is that string operators commute with the generators of the
stabilizer. Also observe that, let us say, a red plaquette operator
can be identified both with a green string or with a blue string,
see Fig.~\ref{borde_barion}. In both cases the paths are boundaries,
but in the first case it is a boundary for the green shrunk lattice
and in the second for the blue one.

We can now relate $\Z_2$ homology theory \cite{optimalGraphs} and
string operators. We recall that a closed path is a boundary iff it
is a combination of boundaries of plaquettes. For the, say, red shrunk
lattice, green and blue plaquettes make up the set of its plaquettes. Thus,
two string operators of the same color are equivalent iff their
corresponding paths are homologous, that is, if they differ by a
boundary. Then it makes sense to label the string operators as
$S_\mu^{C\sigma}$, where $C$ is a color, $\sigma$ is a Pauli matrix
and $\mu$ is a label related to the homology class. But what about
the equivalence of strings of different colors?
Fig.~\ref{toro_color} shows how the product of a pair of homologous
red and blue strings related to the same Pauli matrix produces a
green string. Note that at those qubits in which both strings cross
they cancel each other. In general we have
\begin{equation}\label{combina_colores}
S_\mu^{R\sigma}S_\mu^{G\sigma}S_\mu^{B\sigma}\sim 1.
\end{equation}
This property gives the interplay between homology and color, as we
will see later.

The commutation properties of strings are essential to their study
as operators on $\mathcal C$. It turns out that:
\begin{equation}\label{commutadores}
[S_\mu^{C\sigma},S_\nu^{C'\sigma}]=
[S_\mu^{C\sigma},S_\mu^{C'\sigma'}]=
[S_\mu^{C\sigma},S_\nu^{C\sigma'}]=0.
\end{equation}
The first commutator is trivially null; for the second, note that
two homologous strings must cross an even number of times; the third
is zero because two strings of the same color always share an even
number of qubits.
 Other commutators will depend on the homology, they will be
non-zero iff the labels of the strings are completely different and
closed paths in the respective homology classes cross an odd number
of times. For example, consider the torus with the labels $1$ and
$2$ for its two fundamental cycles. If we make the identifications
\begin{alignat}4\label{}
Z_1 &\leftrightarrow S_1^{R Z}, Z_2 &\leftrightarrow S_1^{G Z},
Z_3 &\leftrightarrow S_2^{R Z}, Z_4 &\leftrightarrow S_2^{G Z}, \\
X_1 &\leftrightarrow S_2^{G X}, X_2 &\leftrightarrow S_2^{R X}, X_3
&\leftrightarrow S_1^{G X}, X_4 &\leftrightarrow S_1^{R X},
\end{alignat}
then we recover the usual commutation relations for Pauli operators
in $\mathcal H_2^4$.

\begin{figure}
 \includegraphics[width=8 cm]{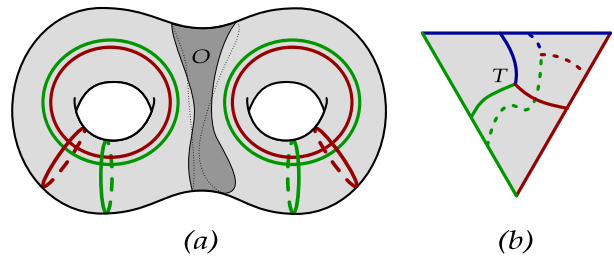}
 \caption{(a) The grey area is the support of an operator $O$ in $\mathcal Z$.
 It must be trivial since it commutes with the colored string operators shown,
 which are enough to construct all $X$ and $Z$ operators for logical qubits.
  (b) The color structure of a planar triangular code. A 3-string operator $T$ and
   a deformation of it are displayed, showing why $\sset{T^X,T^Z}=0$.}
  \label{operadores_triviales}
\end{figure}

We now determine the distance of
 color codes. Recall that in
order to calculate the distance we must find the smallest length
among those operators in $\mathcal Z$ which act nontrivially on
$\mathcal C$. Let the support of an operator in $\mathcal Z$ be the
set of qubits in which it acts nontrivially. We can identify this
support with a set of sites in the lattice. The point is that any
operator in $\mathcal Z$ such that its support does not contain a
closed path which is not a boundary, must be in $\mathcal S$. The
idea behind this assertion is illustrated in
Fig.~\ref{operadores_triviales} . For such an operator $O$, we can
construct a set of string operators with two properties: their
support does not intersect the support of $O$ and any operator in
$\mathcal S$ commuting with all of them must be trivial. The
distance thus is the minimal length among paths with nontrivial
homology.

\begin{figure}
 \includegraphics[width=8cm]{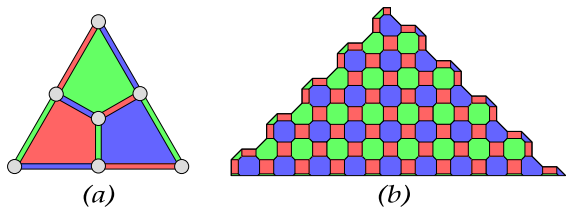}
 \caption{
 (a) The simplest example of a triangular code. The original lattice
 in the sphere can be recovered by adding a site and linking it to
 the sites at vertices of the triangle.
 (b) Triangular codes of any size can be constructed with the special property that any
 plaquette has $v=4m$ sites, with $m$ an integer. This extra requirement is
needed in order to implement the phase-shift gate $K$.
  \label{figura_triangulos}
}
  \label{octogonales}
\end{figure}

Strings are all we need to handle tori of arbitrary genus.
Things get more interesting if we consider oriented surfaces with
border, which can be obtained by opening holes in a closed surface.
In particular, we will introduce holes by removing plaquettes. If we
remove, for example, a green plaquette, green strings can have an
endpoint on it, but not blue or red ones. Then borders have a color,
and only a green string can end at a green border, see
Fig.~\ref{borde_barion}. The most important case of such bordered
codes are \emph{triangular codes}. They are constructed starting
with a color code in a sphere from which a site and its neighboring
three links and three plaquettes are removed. From constraints
\eqref{operadores_sobrantes} we observe that two generators of the
stabilizer are removed in the process. Since a color code in the
sphere encodes zero qubits, a triangular code will encode a single
qubit. Examples of triangular codes are displayed in
Fig~\ref{figura_triangulos}.

So let us show why new features are introduced through triangular
codes. Observe that equation \eqref{combina_colores} suggests the
construction displayed in Fig.~\ref{borde_barion}: three strings,
one of each color, can be combined at a point and obtain an operator
that commutes with plaquette operators.
Fig.~\ref{operadores_triviales}(b) shows the color structure of the
borders in a triangular code. Let $T^\sigma$, $\sigma\in\sset{X,Z}$,
be the 3-string operators depicted in the figure. By deforming $T$ a
little it becomes clear that $\sset{T^X,T^Z}=0$, because $T$ and its
deformation cross each other once at strings of different colors.
Such an anticommutation property is impossible with strings because
of \eqref{commutadores}.

Although 3-string operators can be used to construct an operator
basis for the encoded qubit in a triangular code, this can
equivalently be done with the operators $\hat X$ and $\hat Z$. They
commute with the stabilizer operators and $\sset{\hat X, \hat Z}=0$
because the total number of qubits is odd.
The generators of the Clifford group are the Hadamard gate $H$ and
the phase-shift gate $K$ applied to any qubit and the controlled-not
gate $\Lambda (X)$ applied to any pair of qubits:
\begin{equation}\label{}
H = \frac 1 {\sqrt 2}\begin{pmatrix} 1 & 1 \\ 1 & -1\end{pmatrix}, %
K = \begin{pmatrix} 1 & 0 \\ 0 & i\end{pmatrix}, %
\Lambda(X) = \begin{pmatrix} I_2 & 0 \\ 0 & X\end{pmatrix}.
\end{equation}
The action of these gates is completely determined up to a global
phase by their action on the operators $X$ and $Z$ of individual
qubits, for example
\begin{equation}\label{}
H^\dagger X H = Z,\quad H^\dagger Z H = X.
\end{equation}
Now consider $\hat H$, $\hat K$ and $\hat \Lambda(X)$. Of course,
$\hat \Lambda(X)$ acts pairwise on two code layers that must be
placed one on top of the other so that the operation is locally
performed. The fact is that in the triangular codes both $\hat H$
and $\hat \Lambda(X)$ act as the local ones at the logical level,
for example:
\begin{equation}\label{}
\hat H^\dagger \hat X \hat H = \hat Z,\quad \hat H^\dagger \hat
Z\hat H = \hat X.
\end{equation}
Unfortunately, $\hat K$ is more tricky because in general it does
not take ground states to ground states. This is so because $\hat K
B_p^X \hat K^\dagger = (-1)^{v/2}B_p^X B_p^Z$ if the plaquette $p$
has $v$ sites.
 However, this difficulty can be overcome by choosing
a suitable lattice, as shown in Fig.~\ref{figura_triangulos}.
For such a suitable code, if the number of sites is congruent with 3
mod 4, then $\hat K$ acts like $K^\dagger$, but this is a minor
detail.
As a result, any operation in the Clifford group can be performed on
certain triangular codes in a fault tolerant way and without
selective addressing. As for the distance of triangular codes, it
can be arbitrarily large: notice that an operator in $\mathcal Z$
acting nontrivially on $\mathcal C$ must have a support connecting
the red, green and blue borders.

 We can give an expression for the states of the logical
qubit $\sset{\ket {\bar 0},\ket {\bar 1}}$:
\begin{equation}\label{}
\ket {\bar 0} := 2^{(1-n)/ 2} \prod_b (1+B_b^X) \prod_p (1+B_p^X)
\,\, \ket 0^{\otimes n}
\end{equation}
and $\ket {\bar 1} := \hat X \ket {\bar 0}$, so that $\hat Z \ket
{\bar l}=(-1)^l\ket {\bar l}$, $l = 0,1$. Observe that if we have a
state in $\mathcal C$ and we measure each physical qubit in the $Z$
basis we are also performing a destructive measurement in the $\hat
Z$ basis. This is so because the two sets of outputs do not have
common elements. In fact, the classical distance between any output
of $\ket {\bar 0}$ and any of $\ket {\bar 1}$ is at least $2t+1$.
Moreover, we can admit faulty measurements, since the faulty
measurement of a qubit is equivalent to an $X$ error previous to it.
In this sense, the measuring process is as robust as the code
itself.

 Now let us return to the general case of an arbitrary color
code in a surface with border. We can give a Hamiltonian such that
its ground state is $\mathcal C$:
\begin{equation}\label{Hamiltoniano}
H= - \sum_p B_p^X - \sum_p B_p^Z.
\end{equation}
Observe that color plays no role in the Hamiltonian, rather, it is
just a tool we introduce to analyze it. Any eigenstate $\ket \psi$
of $H$ for which any of the conditions $B_p^\sigma \ket \psi =
\ket \psi$ is not fulfilled will be an excited state. Then we can
say, for example, that an state $\ket \psi$ for which $B_p^X \ket
\psi = - \ket \psi$ has an $X$-type excitation or quasi-particle at
plaquette $p$. These excitations have the color of the plaquette
where they live.
 In a quantum system with this hamiltonian and the
geometry of the corresponding surface, any local error will either
leave the ground state untouched or produce some quasiparticles that
will decay.
This family of quantum systems shows topological quantum order:
they become self-protected from local errors by the gap
\cite{kitaevpreskill06}, \cite{levinwen06}.

As a final remark, we want to point out that the ability to perform
fault tolerantly any operation in the Clifford group is enough for
universal quantum computation as long as a reservoir of certain
states is available \cite{bravyikitaev05}. These states need not be
pure, and so they could be obtained, for example, by faulty methods,
perhaps semi-topological ones. Namely, one can distill these
imperfect states until certain magic states are obtained
\cite{bravyikitaev05}. These magic states are enough to perform
universal quantum computation with the Clifford group, which is
different from topological computation based on
braiding quasiparticles \cite{kitaev97}, \cite{freedman_etal01},
\cite{preskillnotes}.

\noindent {\em Acknowledgements} H.B. acknowledges useful discussions with
L. Tarruell.
We acknowledge financial support
from a PFI fellowship of the EJ-GV (H.B.), DGS grant  under contract
BFM 2003-05316-C02-01 (M.A.MD.), and CAM-UCM grant under ref.
910758.

\end{document}